\documentclass[lettersize,journal]{IEEEtran}
\usepackage{amsmath,amsfonts}
\usepackage[ruled,linesnumbered]{algorithm2e}
\usepackage{array}
\usepackage{textcomp}
\usepackage{stfloats}
\usepackage{url}
\usepackage{verbatim}
\usepackage{graphicx}
\usepackage{cite}
\usepackage{multirow}
\usepackage{booktabs}
\usepackage{soul}
\usepackage{subcaption}  
\usepackage[normalem]{ulem}
\usepackage{pifont,xcolor}
\usepackage{xfrac}
\useunder{\uline}{\ul}{}

\begin{document}

\title{Treat Traffic Like Trees: A Semantic-Preserving Hierarchical Graph-Based Expert Framework for Encrypted Traffic Analysis}




\author{Yuantu Luo, Jun Tao\textsuperscript{\dag}, Linxiao Yu, Guang Cheng,~\IEEEmembership{Member,~IEEE}
\thanks{
   The authors are with the School of Cyber Science and Engineering, Southeast University, Nanjing 211189, China, 
   also with Purple Mountain Laboratories, Nanjing 210096, China, 
   also with Engineering Research Center of Blockchain Application, Supervision and Management (Southeast University), MoE, Nanjing 211189, Jiangsu, China, 
   and also with the Jiangsu Province Engineering Research Center of Security for Ubiquitous Network, Nanjing 211806, China (e-mail: \{ytluo, juntao, yulinxiaoybbb, chengguang\}@seu.edu.cn).
}
\thanks{\textsuperscript{\dag}Corresponding author.}
\thanks{\textbf{This work has been submitted to the IEEE for possible publication. 
Copyright may be transferred without notice, after which this version may no longer be accessible. }}
}

\markboth{Journal of \LaTeX\ Class Files,~Vol.~14, No.~8, August~2021}%
{Shell \MakeLowercase{\textit{et al.}}: A Sample Article Using IEEEtran.cls for IEEE Journals}

\maketitle

\begin{abstract}

Graph-based deep learning methods have been widely employed in encrypted traffic analysis to exploit latent correlations across different granularities. 
However, while complex preprocessing pipelines and sophisticated model structures often achieve strong performance, they may obscure inherent protocol semantics during representation learning. 
Moreover, the hierarchical structure of protocol layers and their corresponding fields, defined by protocol specifications and routinely utilized in manual traffic analysis, remains underexplored in existing learning frameworks. 
In this paper, we propose Protocol Tree Graph Attention with Mixture of Experts (PTGAMoE), a semantic-preserving hierarchical graph-based expert framework for encrypted traffic analysis. 
The field-based graph construction and expert committee design enable PTGAMoE to quantify the model's preferences for specific fields and protocols. 
Extensive experimental results on representative benchmark datasets under strict no-data-leakage settings demonstrate that PTGAMoE significantly outperforms state-of-the-art (SOTA) models. 
Furthermore, the semantic-preserving design provides interpretable insights into protocol-level feature importance and expert-level contributions, reflecting the model's decision-making logic in encrypted traffic classification tasks. 

\end{abstract}


\section{Introduction}

\IEEEPARstart{E}{ncrypted} traffic has become ubiquitous across various Internet communication scenarios. 
Protocols such as TLS, DTLS, and QUIC are commonly used to encapsulate data payloads in applications ranging from web browsing and media streaming to real-time interactions \cite{backgroud_survey1}. 
Furthermore, the widespread deployment of the TLS 1.3 specification has significantly enhanced data privacy, as its advanced cipher suites and secure encapsulation mechanisms mitigate information leakage risks \cite{Intro_TLS1}\cite{Intro_TLS2}. 
However, the increasing prevalence of encryption poses significant challenges for network management, as payload-obfuscated packets are inherently difficult to identify and classify. 

Machine Learning (ML) and Deep Learning (DL) have demonstrated remarkable success in domains such as Natural Language Processing (NLP) and Computer Vision (CV), prompting significant interest in their application to encrypted traffic analysis. 
Nevertheless, most supervised learning frameworks require input data to be structured as an $N \times D$ dataset $S$, where $N$ represents the number of entries and $D$ denotes a fixed dimension for each entry. 
This rigid constraint often leads to a fundamental semantic mismatch and introduces significant noise into the representation learning process. 

\begin{figure}[htbp]
    \centering
    \includegraphics[width=\linewidth]{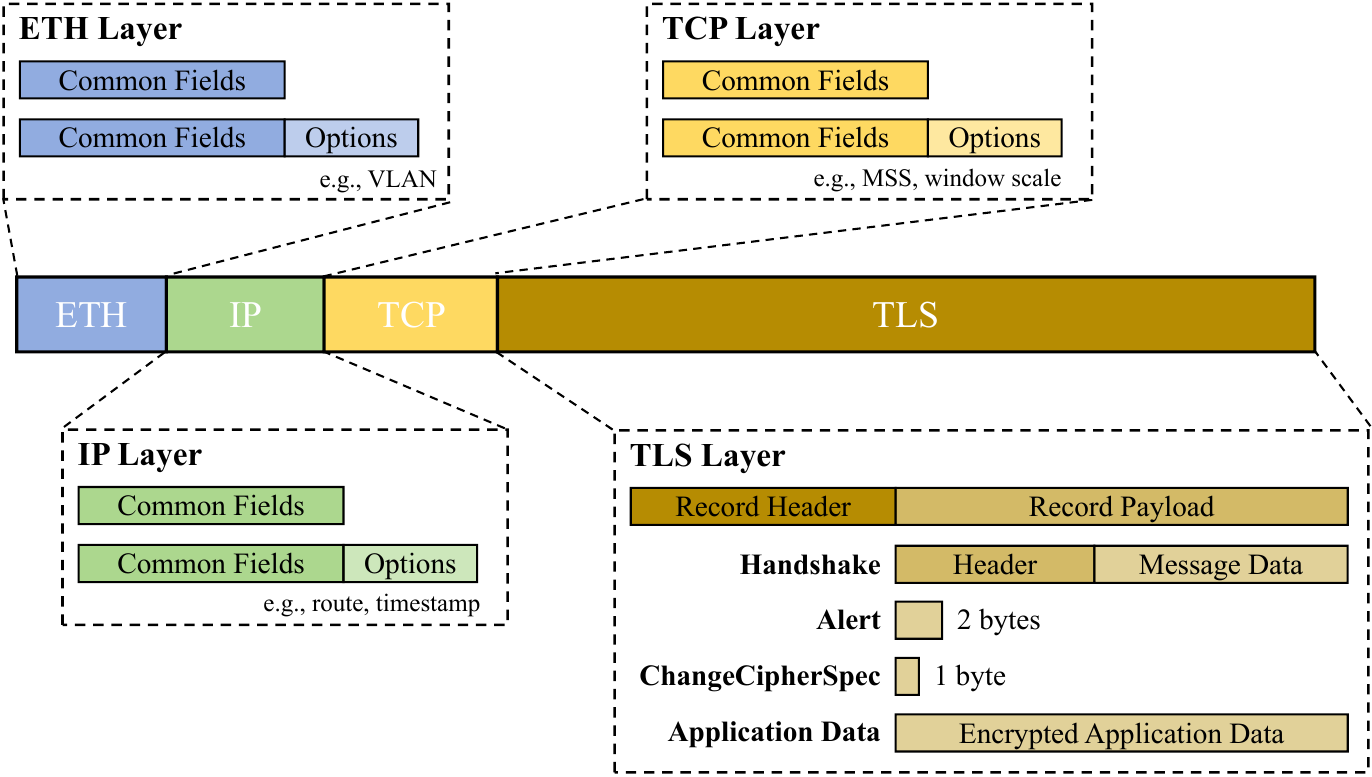}
    \caption{Hierarchical field structure of a typical TCP-based encrypted packet conversation.}
    \label{fig: field_structure}
\end{figure}

As illustrated in Fig.~\ref{fig: field_structure}, TCP-based encrypted packets are constructed according to a layered protocol stack governed by rigorous standards and RFC specifications. 
While a complete stack typically comprises Ethernet (ETH), Internet Protocol (IP), Transport Control Protocol (TCP), and Transport Layer Security (TLS) layers, not all packets within a flow possess a uniform structure \cite{RFC_TLS1.3}. 
Depending on their specific function, the composition of layers and fields varies significantly. 
For instance, session maintenance packets (e.g., keep-alives) may only reach the TCP layer, congestion control mechanisms often introduce variable TCP options, and TLS packets exhibit distinct headers depending on whether they are performing a handshake or transmitting application data. 
Consequently, traditional preprocessing techniques that rely on padding or truncation to force these heterogeneous packets into fixed-shape tensors inevitably disrupt inherent protocol semantics and introduce artificial artifacts. 

To address these issues, researchers have explored various techniques, ranging from statistical feature engineering to sophisticated DL model architectures. 
Statistical attributes, such as flow duration, protocol distributions, and temporal metrics, are extracted to comprehensively capture correlations between traffic flows. 
Beyond these, graph-based representation learning, particularly Graph Neural Networks (GNNs), has been introduced to exploit latent dependencies within traffic data \cite{RW_gnn3_GSPB}\cite{RW_gnn4_AGMF}. 
Due to the inherent correlation between nodes and edges, GNNs are widely employed to model interactions in packets and flows \cite{Method_GAT}\cite{Method_GCN-RTG}\cite{RW_gnn5_flow}. 


However, viewing traffic solely through these feature sets often overlooks the structural nature of layered packets. 
In practice, traffic analyzers like Wireshark employ dissectors to systematically parse raw data into a hierarchical, tree-like representation termed a Dissection Tree (DT)\cite{dissection_tree}. 
Starting from the outermost frame, dissectors recursively expand each encapsulated protocol into its constituent fields. 
Furthermore, individual fields are often broken down into meaningful subfields, such as specific flags or status codes, revealing multiple layers of semantic detail. 
Therefore, the resulting structure is an explicit manifestation of a protocol's syntax and logic, capturing the precise, context-dependent meaning of every byte. 
Crucially, this tree-like representation is a graph by definition, making it naturally suited for analysis with GNNs. 


Additionally, the Mixture of Experts (MoE) architecture has demonstrated significant success within Large Language Models (LLMs) \cite{MoE5_survey}. 
Beyond its training and inference efficiency, MoE is particularly adept at handling disparate data distributions and multi-modal inputs. 
A gating network dynamically routes inputs to the most relevant experts, intelligently synthesizing their outputs to produce a final, weighted decision. 
These two core characteristics, i.e., data specialization and adaptive aggregation, make MoE a highly promising approach for tackling the heterogeneous data structures found in encrypted traffic. 

In this paper, we propose \emph{Protocol Tree Graph Attention with Mixture of Experts} (PTGAMoE). 
Instead of modeling traffic flows as flat byte sequences or fixed-length feature vectors, PTGAMoE explicitly exploits the protocol parsing structure by representing packet fields as protocol tree graphs that mirror real-world encapsulation semantics. 
Layer-specific graph attention experts are employed to capture heterogeneous protocol characteristics, and a MoE fusion module adaptively fuses multi-layer representations with optional flow-level statistics. 
Finally, a permutation-invariant aggregation mechanism is employed to distill packet-level representations into unified flow descriptors, facilitating robust classification in strict, flow-isolated scenarios. 
Furthermore, type-aware field embedding and hierarchical gating mechanisms are incorporated to enhance semantic fidelity and interpretability. 

Our main contributions can be summarized as follows:
\begin{itemize}
    \item \textbf{Protocol Tree Graph Attention (PTGA):} We propose a semantic-preserving graph module that explicitly models hierarchical dependencies between protocol fields, ensuring structural integrity without disruptive padding or truncation.
    \item \textbf{Protocol-Aware MoE Architecture:} We design a layer-wise Mixture of Experts framework that aligns specialized experts with distinct protocol layers, enabling adaptive semantic fusion across heterogeneous and encapsulated protocol structures. 
    \item \textbf{Permutation-Invariant Flow Aggregation:} We introduce a robust aggregation mechanism to distill packet-level representations into unified flow descriptors, facilitating accurate classification under strict flow-isolated deployment scenarios. 
    \item \textbf{Strict No-Leakage Evaluation:} We validate our framework on modern TLS1.3 datasets under rigorous no-data-leakage settings. Results show that PTGAMoE significantly outperforms SOTA models (e.g., ET-BERT, YaTC, RBLJAN) while offering quantifiable interpretability via proposed NGI and GCR metrics.
\end{itemize}

\section{Related Works} \label{sec: related_works}

\subsection{Representation Learning for Encrypted Traffic}

Encrypted traffic analysis has evolved from manual feature engineering to deep representation learning. 
Early works like Kitsune~\cite{RW_representation1_Kitsune} and FS-Net~\cite{RW_representation7_FS-NET} utilized ensembles of autoencoders and recurrent networks to capture flow-level statistics and sequential patterns. 
With the success of pre-training paradigms, models such as ET-BERT~\cite{RW_representation3_ETBERT}, YaTC~\cite{RW_representation4_Yet}, and the recent TrafficFormer~\cite{RW_representation2_TrafficFormer} have leveraged Transformers and Masked Autoencoders to learn contextualized datagram representations from large-scale unlabeled data. 
To improve robustness, data augmentation~\cite{RW_representation6_NetTraceAug} and semantic analysis of packet patterns~\cite{RW_representation5_PktSemantics} have also been explored. 
Furthermore, context learning has been applied to detect sophisticated DPI evasion~\cite{RW_representation8_YouDPI}, while adversarial studies like TANTRA~\cite{RW_representation9_TimeTrantra} highlight the vulnerability of timing-based features. 
RBLJAN~\cite{sota-tdsc25-xiao} comprises a classifier and an adversarial traffic generator at both packet-level and flow-level to capture implicit correlations
between bytes and labels, enabling the construction of powerful packet representations. 
Despite these advances, existing methods predominantly treat traffic as flat byte sequences or statistical vectors, which overlooks the intrinsic inter-relationships between protocol fields, leading to a loss of protocol semantics and limited interpretability. 

\subsection{Graph-Based Modeling in Network Traffic Analysis}

To capture structural dependencies, researchers have increasingly adopted Graph Neural Networks (GNNs). 
For instance, DGNN~\cite{RW_gnn1_DGNN} constructs interaction graphs to identify darknet applications, while FlowGNN~\cite{RW_gnn2_FlowGNN} models encrypted traffic by exploiting relationships between packets within a flow. 
Furthermore, DigTraffic~\cite{sota-aaai25-qiu} further introduces heterogeneous edge designs and graph transformers to represent flow-level message interactions. 
While these graph-based approaches provide a more flexible relational inductive bias than sequential models, they predominantly focus on inter-packet or inter-flow relationships. 
The internal hierarchical parsing structure of individual packets, as revealed by protocol dissectors, is rarely used as the primary modeling target. 
In other words, while graphs are employed to model traffic entities, the protocol-level tree structure that governs field dependencies and encapsulation semantics is not explicitly encoded. 
This leaves a gap between graph-based learning and the intrinsic hierarchical organization of network protocols. 

\subsection{Mixture of Experts for Heterogeneous Traffic Semantics}

The Mixture of Experts (MoE) architecture has become a cornerstone of contemporary Large Language Models (LLMs)~\cite{MoE1_KIMI, MoE2_Google, MoE3_Xiaomi, MoE4_Qwen}, demonstrating an exceptional capacity to process heterogeneous data by selectively activating specialized sub-networks~\cite{RW_moe2_SwitchTransformer}. 
This paradigm has recently extended to time-series foundation models~\cite{RW_moe3_Time-MoE}\cite{RW_moe4_Moirai-MoE} and emerging network traffic foundation models. 
Within the encrypted traffic classification domain, MoE has shown significant promise. 
For instance, CL-ViME~\cite{RW_moe1_CL-ViME} integrates MoE with contrastive learning to facilitate dual-view feature extraction. 
However, existing traffic-oriented MoE frameworks typically perform expert routing over flattened sequences or vision-mapped representations, rather than aligning experts with functional protocol layers. 
Given that network traffic is inherently structured into hierarchical layers with heterogeneous specifications, the protocol stack provides a natural substrate for layer-wise expert specialization. 
This observation motivates our PTGAMoE framework, which utilizes a dedicated expert committee to model the natural hierarchy of network protocols while preserving structural integrity. 

\subsection{Strict Scenarios in Traffic Classification Tasks}

Recently, the research community has voiced significant concerns regarding a credibility crisis in encrypted network traffic classification. 
Studies indicate that the astonishing performance reported in much of the literature, often exceeding 98\% accuracy, is frequently inflated by methodological pitfalls. 
Specifically, models often become shortcut learners by exploiting Strong Identification Information (SII), such as IP/MAC addresses and port numbers, which act as uninformative artifacts that prevent genuine generalization to real-world scenarios~\cite{RW_sok_enigma}. 
Furthermore, traditional per-packet dataset splitting introduces severe data leakage, as session-specific implicit identifiers allow models to link test packets to training labels~\cite{RW_sok_sweet}. 
To address these flaws, researchers advocate for the adoption of \textbf{Strict Scenarios}, which necessitate flow-isolated splitting and the exclusion of SII to rigorously evaluate a model’s ability to learn behavior-driven protocol semantics rather than superficial shortcuts. 

\section{Overview}

\begin{figure}[htbp]
    \centering
    \includegraphics[width=\linewidth]{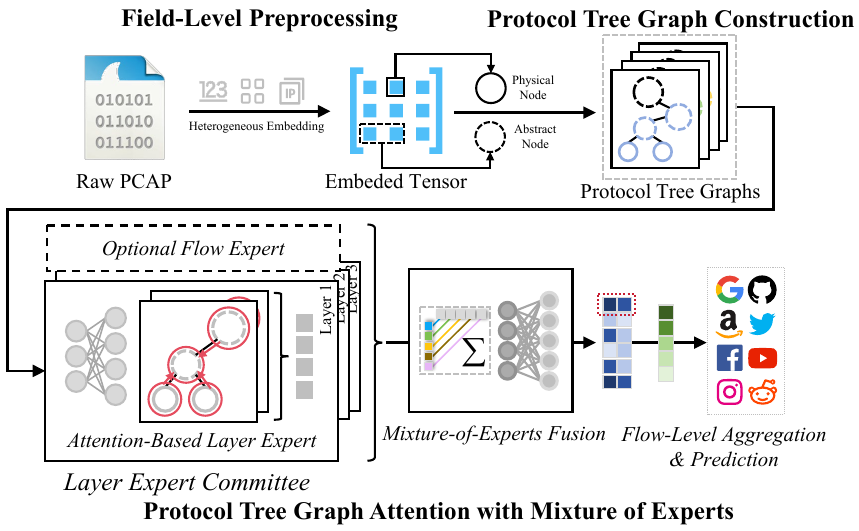}
    \caption{Overview of the PTGAMoE workflow. }
    \label{fig: overview}
\end{figure}

We propose Protocol Tree Graph Attention with Mixture of Experts (PTGAMoE), a structure-aware learning framework designed to preserve the semantic organization of network traffic and enhance the interpretability of protocol-level feature contributions. 
As illustrated in Fig.~\ref{fig: overview}, the PTGAMoE workflow consists of three main stages: Field-Level Preprocessing, Protocol Tree Graph Construction, and Protocol Tree Graph Attention with Mixture of Experts architecture. 

\noindent\textbf{Field-Level Preprocessing.} 
The workflow begins with field-level preprocessing, which transforms raw PCAP files into numerical tensors suitable for representation learning. 
Instead of treating traffic data as flat byte sequences or fixed-length vectors, PTGAMoE performs semantics-aware field embedding, where each traffic field is embedded according to its inherent type and protocol meaning. 
Specifically, address fields, numerical fields, and categorical fields are processed via dedicated embedding strategies, yielding a unified tensor that maintains protocol-level semantics while ensuring compatibility with downstream graph modeling. 

\noindent\textbf{Protocol Tree Graph Construction.} 
Based on the embedded field representations, PTGAMoE constructs a Protocol Tree Graph (PTG) that explicitly models the hierarchical structure of protocol formats. 
In the PTG, traffic fields are mapped to graph nodes, and edges are defined according to the parent–child relationships inherent in protocol specifications. 
Beyond these hierarchical dependencies, we introduce original structural components, including a set of abstract nodes and a global summarizer, to explicitly encode structural protocol identities and facilitate layer-wide semantic exchange. 
This graph abstraction enables PTGAMoE to represent traffic packets as structured objects rather than flat feature collections, forming the structural foundation for subsequent graph-based learning. 

\noindent\textbf{Protocol Tree Graph Attention with Mixture of Experts.} 
Given the constructed PTGs, PTGAMoE employs a layer-wise mixture-of-experts architecture to perform structure-aware representation learning. 
Each protocol layer is associated with an attention-based graph expert that operates on its corresponding protocol tree graph, enabling the model to capture layer-specific semantics while mitigating noise introduced by heterogeneous packet structures. 
An optional flow-level expert can be incorporated to leverage aggregated flow statistics when available. 
The outputs of all experts are adaptively fused through learnable gating mechanisms to generate robust packet-level representations. 
Furthermore, a permutation-invariant aggregation mechanism is utilized to distill these packet-level signals into a unified flow descriptor, facilitating accurate classification in strict, flow-isolated scenarios. 
Finally, the resulting flow representation is passed to a prediction head for final classification. 

\section{Field-Level Preprocessing} \label{sec: preprocessing}

This section illustrates the whole procedure of field-level preprocessing, including streaming field extraction and heterogeneous field embedding. 
These two steps convert raw traffic data into tensors, which can be the standard input in the following models. 

The preprocessing stage transforms raw PCAP files into numerical tensors suitable for graph representation learning. 
First, raw traffic is parsed into structured field-level attributes using a streaming PCAP-to-CSV converter. 
To ensure scalability and bounded memory consumption in high-concurrency scenarios, this converter utilizes a chunk-wise parsing strategy (see Appendix~\ref{apdx: streaming_field_extraction}). 

\begin{figure}[htbp]
    \centering
    \includegraphics[width=\linewidth]{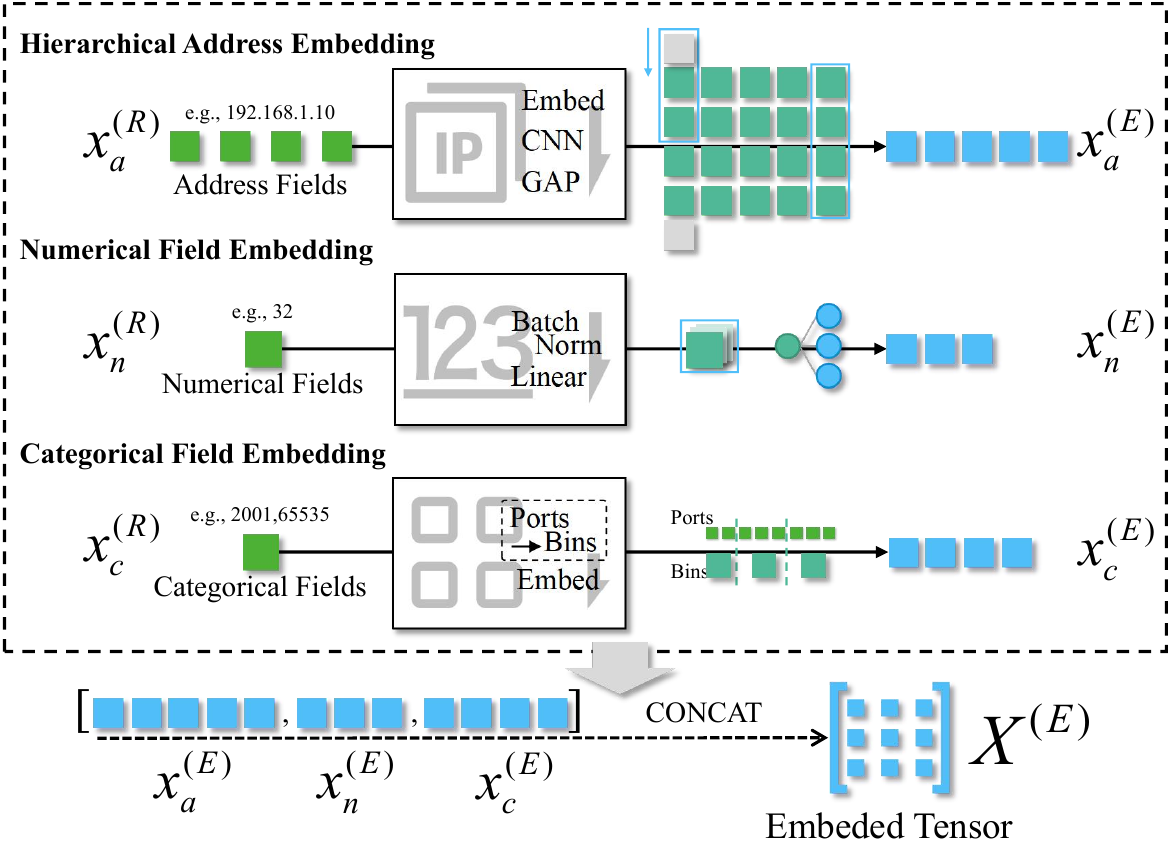}
    \caption{The field-level embedding procedure. }
    \label{fig: embedding}
\end{figure}

Following field extraction from raw PCAPs, the subsequent step is field embedding, which transforms discrete hexadecimal or decimal values into dense vector representations suitable for deep learning models. 
To address the multi-modal nature of network traffic, including comprising addresses, continuous values and discrete codes, we design a unified embedding module that projects these disparate field types into a shared high-dimensional latent space. 

As illustrated in Fig.\ref{fig: embedding}, let $X^{(R)}$ denote an raw CSV input packet, which is decomposed into three subsets, comprising address fields $x^{(R)}_{a}$, numerical fields $x^{(R)}_{n}$ and categorical fields $x^{(R)}_{c}$. 
Consequently, the embedding module includes three specialized sub-components tailored to these distinct data types. 
The final output is an embedded tensor $X^{(E)}$, containing the vector representations $\mathbf{x}^{(E)}_{a}$, $\mathbf{x}^{(E)}_{n}$, and $x^{(E)}_{c}$ corresponding to their respective inputs. 
Formally, each traffic field is mapped to a fixed-dimensional vector through a type-specific embedding function 
\begin{equation}
    \mathbf{x}^{(E)}_{t} = f_{t} \left( x^{(R)}_{t} \right), \quad t \in \{a, n, c\}, 
\end{equation}
where $f_{t}(\cdot)$ denotes the embedding function for address, numerical, or categorical fields, respectively. 

\paragraph{Hierarchical Address Embedding (HAE)} 
The HAE module is specifically constructed for IP and MAC addresses. 
These fields possess a unique hierarchical structure where semantics are encoded in byte-level segments (e.g., subnet masks in IPv4, Organizationally Unique Identifiers in MAC). 
Treating them as monolithic categorical strings would ignore this intrinsic structure and result in an intractable vocabulary size. 
To capture this structure, we split an address field $x^{(R)}_{a}$ into a sequence of octets $(o_{1}, o_{2}, \dots, o_{K})$, where $K=4$ for IPv4 and $K=6$ for MAC. 
Each octet is independently embedded, and the resulting sequence is aggregated using a 1D Convolutional Neural Network (CNN). 
This design allows the model to learn local dependencies between adjacent bytes (e.g., network prefixes). 
Finally, Global Average Pooling (GAP) is applied to obtain a fixed-length representation. 
The procedure is formalized as 
\begin{align}
    \mathbf{s}_a &= \phi_{\text{oct}}(x^{(R)}_a) = (o_1, o_2, \dots, o_K), \\
    \mathbf{x}^{(E)}_{a} &= f_a(\mathbf{s}_a) = \operatorname{GAP}\!\left(\operatorname{CNN}_{1\mathrm{D}}\left(\operatorname{Embed}(\mathbf{s}_a)\right)\right),  
\end{align}
where $\phi_{\text{oct}}(\cdot)$ denotes a deterministic transformation
that decomposes an address into a sequence of byte-level octets according
to protocol specifications. 
Notably, $\operatorname{Embed}(\cdot)$ denotes a generic learnable
embedding lookup for discrete symbols, whose parameters are jointly
optimized during training. 

\paragraph{Numerical Field Embedding}
This module handles continuous numerical fields (e.g., packet length, window size). 
Unlike categorical features, these values possess ordinal magnitude semantics. 
To project them into the target dimension $d$, we employ a linear transformation block. 
To ensure numerical stability and accelerate model convergence, Batch Normalization (BatchNorm) is applied prior to the projection:
\begin{equation}
    \mathbf{x}^{(E)}_{n} = \operatorname{Linear}\left(\operatorname{BatchNorm}\left(x^{(R)}_{n}\right)\right). 
\end{equation}

\paragraph{Categorical Field Embedding}
Categorical field embedding is applied to discrete protocol fields with finite or discretized vocabularies, such as protocol flags and port numbers. 
For fields with extremely large or sparse vocabularies (e.g., raw port numbers), we first employ a binning strategy that maps raw values to a smaller set of semantically meaningful bins. 
This step reduces vocabulary size, alleviates sparsity, and improves representation robustness. 
After discretization, categorical values are mapped to dense vectors
through a learnable embedding function: 
\begin{equation}
    \mathbf{x}^{(E)}_{c} = \operatorname{Embed}\left(x^{(R)}_{c}\right). 
\end{equation}

The final embedded representations of all fields are concatenated into a packed tensor 
\begin{equation} \label{eq:embedded_tensor}
    \mathbf{X}^{(E)} = \operatorname{CONCAT}\left( \mathbf{x}^{(E)}_{a}, \mathbf{x}^{(E)}_{n}, \mathbf{x}^{(E)}_{c} \right), 
\end{equation}
in which the index of each field's embedding is recorded for the subsequent graph node initialization stage. 

\section{Layer-wise Protocol Tree Graph Representation} 

This section presents the construction of the Protocol Tree Graph (PTG), including its graph representation and the principles used to derive nodes and edges from protocol semantics. 

Due to the layered structure of TCP/IP protocol stack, a network packet can naturally be analyzed in a hierarchical manner. 
Formally, for a specific layer $k$ (e.g., the TLS layer), we define the PTG as an undirected graph 
\begin{equation}
    \mathcal{G}_{k}=\left(\mathcal{V}_{k}, \mathcal{E}_{k}\right), 
\end{equation}
where $\mathcal{V}_{k}$ represents the set of nodes corresponding to protocol fields and $\mathcal{E}_{k}$ denotes the set of edges representing structural dependencies induced by protocol formats. 

\begin{figure}[htbp]
    \centering
    \includegraphics[width=\linewidth]{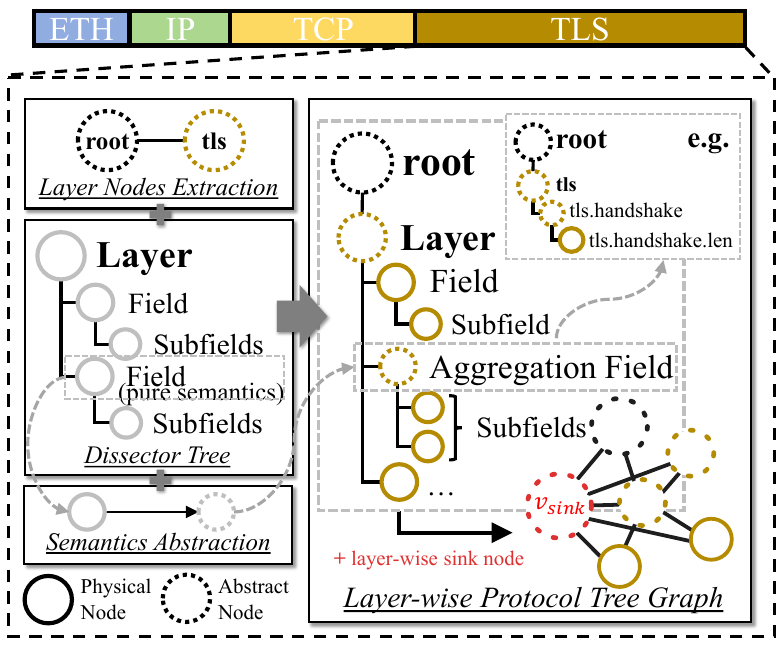}
    \caption{Hierarchical representation of the Protocol Tree Graph (PTG) featuring abstract and physical nodes. }
    \label{fig: PTGrepresentation}
\end{figure}

As illustrated in Fig.~\ref{fig: PTGrepresentation}, the node set $\mathcal{V}_k$ is constructed through layer-level node extraction and parsing of the packet’s Dissection Tree (DT). 
We categorize nodes into two types according to their semantic roles:  
\begin{itemize}
    \item \textbf{Physical Nodes ($\mathcal{V}^{\mathrm{(phy)}}_{k}$)}: Physical nodes correspond to value-carrying protocol fields, which typically appear as leaf nodes in the protocol tree (e.g., \texttt{tls.record.length}, \texttt{tls.handshake.length}). 
    These nodes encode the explicit observable information of network traffic. 
    \item \textbf{Abstract Nodes ($\mathcal{V}^{\mathrm{(abs)}}_{k}$)}: 
    Abstract nodes represent structural protocol components that do not carry explicit field values.
    They include protocol layer nodes $v^{(\mathrm{layer})}_{k}$ (e.g., \texttt{tls}), aggregation fields $v^{(\mathrm{aggr})}_{k}$ (e.g., \texttt{tls.handshake}, \texttt{tls.record}), a global virtual root node ($v^{(\mathrm{root})}_{k}$), and a \textbf{layer-wise sink node} $v^{(\mathrm{sink})}_{k}$. 
    Depending on their position in a layer-wise PTG, abstract nodes may aggregate information from descendant nodes. 
    Notably, $v^{(\mathrm{sink})}_{k}$ is designed as a global latent summarizer and an attention aggregation node. 
    It facilitates the capture of holistic layer-specific semantics that transcend the local tree hierarchy, mitigating potential information bottlenecks during deep message passing and ensuring each field node remains context-aware of the entire protocol layer’s state. 
\end{itemize}
Thus, the total node set for protocol $k$ is defined as 
\begin{equation}
    \mathcal{V}_k = \mathcal{V}_k^{\mathrm{(phy)}} \cup \mathcal{V}_k^{\mathrm{(abs)}}. 
\end{equation}

The edge set $\mathcal{E}_k$ is formally defined as the union of hierarchical protocol edges $\mathcal{E}_k^{\mathrm{(hier)}}$ and global sink edges $\mathcal{E}_k^{\mathrm{(sink)}}$: 
\begin{equation}
    \mathcal{E}_k = \mathcal{E}_k^{\mathrm{(hier)}} \cup \mathcal{E}_k^{\mathrm{(sink)}},
\end{equation}
where $\mathcal{E}_k^{\mathrm{(hier)}} = \{(u_{k}, v_{k}) \mid u_{k} \in \mathrm{Children}(v_{k})\}$ preserves the intrinsic parent-child relationships referring to the DT. 
To model long-range dependencies and perform global feature integration, we introduce $\mathcal{E}_{k}^{\mathrm{(sink)}} = \{(v, v^{(\mathrm{sink})}_k) \mid \forall v \in \mathcal{V}_k \setminus \{v^{(\mathrm{sink})}_k\}\}$, establishing a direct information shortcut between every node and the layer-wise sink. 
In practice, each undirected edge is treated as two directed edges, supporting bidirectional propagation for both fine-grained semantic aggregation and global context distribution. 

Algorithm~\ref{alg:ptg-construction} summarizes the construction procedure of layer-wise PTGs from a dissection tree. 
The input dissection tree is denoted as $\mathcal{T} = (\mathcal{N},\mathcal{R})$. 
$\mathcal{N}$ represents the set of field-level and structural nodes extracted from the dissection tree. 
$\mathcal{R}$ encodes parent-child relationships induced by protocol formats. 
The mapping function $\pi(n)$ associates each dissector node $n \in \mathcal{N}$ with its corresponding protocol layer. 
The final output is a set of layer-wise PTGs $\mathcal{G}$, which serves as structured inputs for subsequent representation learning. 

\begin{algorithm}[t]
\caption{Protocol Tree Graph Construction}
\label{alg:ptg-construction}
\KwIn{
Dissection tree $\mathcal{T}=(\mathcal{N},\mathcal{R})$; \\
Protocol layer set $\mathcal{L}$
}
\KwOut{Layer-wise Protocol Tree Graphs $\mathcal{G} = \{\mathcal{G}_k \mid k\in\mathcal{L}\}$}

\ForEach{$k \in \mathcal{L}$}{
    $\mathcal{V}_k \leftarrow \emptyset$, $\mathcal{V}^{(\mathrm{abs})}_{k} \leftarrow \emptyset $, $\mathcal{V}^{(\mathrm{phy})}_{k} \leftarrow \emptyset$\; 
 
    $\mathcal{E}_k \leftarrow \emptyset$, $\mathcal{E}_k^{\mathrm{(hier)}} \leftarrow \emptyset$, $\mathcal{E}_k^{\mathrm{(sink)}} \leftarrow \emptyset$\;

    $\mathcal{N}_k \leftarrow \{ n \mid n \in \mathcal{N} \cap \pi(n)=k \}$\;
    $\mathcal{R}_k \leftarrow \{ (n_p,n_c) \mid (n_p,n_c)\in\mathcal{R} \cap n_p,n_c\in\mathcal{N}_k \}$\;

    $\mathcal{V}^{\mathrm{(abs)}}_{k} \leftarrow \mathcal{V}^{\mathrm{(abs)}}_{k} \cup \{{v^{(\mathrm{root})}_{k}} , {v^{(\mathrm{layer})}_{k}} , {v^{(\mathrm{aggr})}_{k}}, {v^{(\mathrm{sink})}_{k}}\}$\;
    $\mathcal{E}_{k}^{\mathrm{(hier)}} \leftarrow \mathcal{E}_{k}^{\mathrm{(hier)}} \cup \{({v^{(\mathrm{root})}_{k}}, {v^{(\mathrm{layer})}_{k}}) , ({v^{(\mathrm{layer})}_{k}}, {v^{(\mathrm{aggr})}_{k}})\}$\;

    \ForEach{$n \in \mathcal{N}_k$}{
        \If{\textsc{IsPhysical}(n)}{
            $\mathcal{V}^{\mathrm{(phy)}}_{k} \leftarrow \mathcal{V}^{\mathrm{(phy)}}_{k} \cup \{n\}$\;
        }
    }
    $\mathcal{V}_k \leftarrow \mathcal{V}_k^{\mathrm{(phy)}} \cup \mathcal{V}_k^{\mathrm{(abs)}}$\;  

    \ForEach{$(n_p,n_c) \in \mathcal{R}_k$}{
        \If{$n_{p}, n_{c} \in \mathcal{V}_{k}$}{
            $\mathcal{E}_k^{\mathrm{(hier)}} \leftarrow \mathcal{E}_k^{\mathrm{(hier)}} \cup \{(n_{p}, n_{c})\}$\;
        }
    }
    $\mathcal{E}_{k}^{\mathrm{(sink)}} = \{(v, v^{(\mathrm{sink})}_k) \mid \forall v \in \mathcal{V}_k \setminus \{v^{(\mathrm{sink})}_k\}\}$\; 
    $\mathcal{E}_k = \mathcal{E}_k^{\mathrm{(hier)}} \cup \mathcal{E}_k^{\mathrm{(sink)}}$\;
    $\mathcal{G}_{k}=\left(\mathcal{V}_{k}, \mathcal{E}_{k}\right)$\;
}
$\mathcal{G} \leftarrow \{\mathcal{G}_k \mid k\in\mathcal{L}\}$\;
\Return $\mathcal{G}$\;
\end{algorithm}

To distinguish physical and abstract nodes and to fully exploit their respective semantic roles, we adopt a differential node initialization strategy. 
Let $\mathbf{h}_v^{(E)}$ denote the initial representation of node $v$. 
\begin{itemize}
\item For physical nodes $v \in \mathcal{V}^{\mathrm{(phy)}}$, we initialize features by extracting the corresponding vector from the packed tensor $X^{(E)}$ defined in equation (\ref{eq:embedded_tensor}): 
\begin{equation}
    \mathbf{h}_v^{(E)} = \text{Slice}(X^{(E)}, \text{idx}_v), 
\end{equation}
where $\text{idx}_v$ denotes the index mapping derived from the field-level preprocessing stage. 

\item For abstract nodes $v \in \mathcal{V}^{\mathrm{(abs)}}$, we assign each node an independent learnable semantic token $\mathbf{t}_v \in \mathbb{R}^{D}$:
\begin{equation}
    \mathbf{h}_v^{(E)} = \mathbf{t}_v. 
\end{equation}
These node-specific semantic tokens are jointly optimized during training, enabling the model to encode structural protocol identities prior to message passing while maintaining a consistent embedding dimensionality. 

\end{itemize}

\noindent \textbf{Distinctions Between DT and PTG:} 
Although the PTG construction is inspired by the DT used in tools such as Wireshark, the two structures differ fundamentally in both purpose and representation. 
The DT is designed as a protocol parsing structure, aiming to exhaustively decode packet bytes into protocol fields for inspection and debugging. 
As a result, it treats all parsed fields uniformly and does not explicitly distinguish between value-carrying fields and structural protocol components. 
In contrast, PTG is constructed explicitly for representation learning.
It introduces structurally differentiated abstract nodes to capture protocol hierarchy and packet-level structure, while differentiating between fields with concrete values and structural protocol elements. 
This abstraction enables PTG to preserve protocol semantics and provide a global, model-friendly representation that is suitable for graph-based learning. 

\section{Protocol Tree Graph Attention with Mixture of Experts}

\begin{figure*}[htbp]
    \centering
    \includegraphics[width=\linewidth]{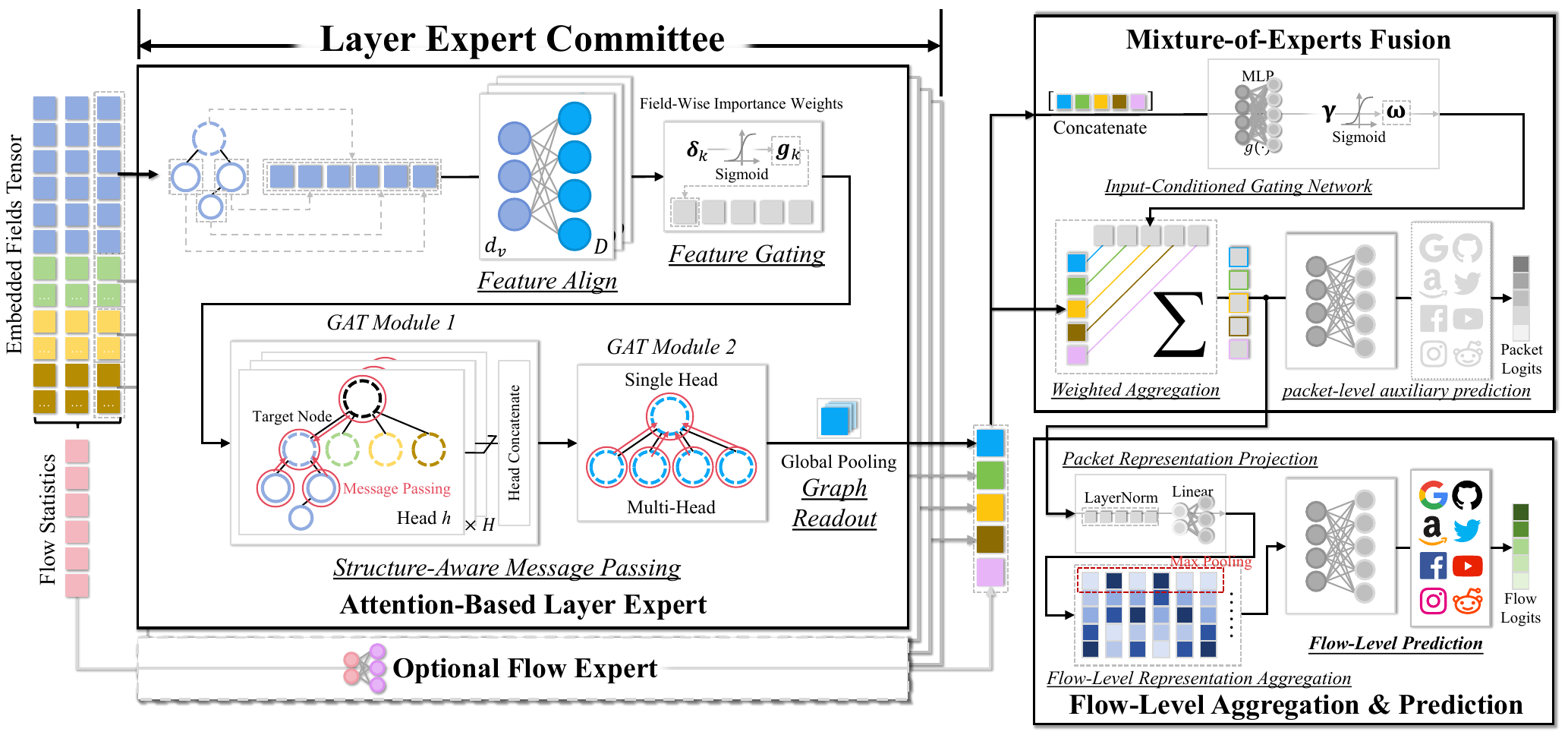}
    \caption{The model design of PTGAMoE. }
    \label{fig: PTGAMoE}
\end{figure*}

This section presents the detailed architecture of our semantic-preserving model for encrypted traffic analysis, namely Protocol Tree Graph Attention with Mixture of Experts (PTGAMoE). 
As illustrated in Fig.~\ref{fig: PTGAMoE}, PTGAMoE consists of four major components: a Layer Expert Committee, an Optional Flow Expert, a Mixture-of-Experts Fusion module, and a Flow-Level Aggregation \& Prediction module.  

The LEC extracts structure-aware representations from protocol tree graphs, with each expert generating an intermediate semantic embedding. 
When enabled, the OFE integrates flow-level statistical features to provide a holistic network view. 
Subsequently, the MoEF module adaptively synthesizes these expert outputs to produce refined packet-level representations. 
Finally, a permutation-invariant mechanism aggregates these signals into a unified flow descriptor for the final classification decision. 

\subsection{Layer Expert Committee}

For a conversation involving $K$ protocol layers, the LEC utilizes $K$ attention-based experts, each specialized for a specific layer-wise PTG. 
Each expert performs structure-aware representation learning through a four-stage pipeline: feature alignment, feature gating, message passing, and graph readout. 

\paragraph{Feature Alignment} 
To enable joint graph-based processing, a feature alignment mechanism is introduced to project all node features into a unified hidden space. 
Let $v \in \mathcal{V}_{k}$ denote a node in the layer PTG of protocol $k$, and let $\mathbf{h}_v^{(E)} \in \mathbb{R}^{d_v}$ be its embedded feature vector obtained from the field-level preprocessing stage in Section~\ref{sec: preprocessing}. 
A linear alignment function is applied to map $\mathbf{h}_v^{(E)}$ to a shared hidden dimension $D$: 
\begin{equation}
    \mathbf{h}_v^{(0)} = \mathbf{W}_{d_v} \mathbf{h}_v^{(E)} + \mathbf{b}_{d_v}, 
\end{equation}
where $\mathbf{W}_{d_v} \in \mathbb{R}^{D \times d_v}$ and $\mathbf{b}_{d_v}$ are learnable parameters associated with the corresponding input dimensionality. 
This design allows fields with different semantic types to be jointly processed while preserving their original representations. 

\paragraph{Feature Gating}
To suppress protocol noise and enhance interpretability, we employ a node-level gating mechanism. 
Specifically, for a layer-wise PTG with $|\mathcal{V}_{k}|$ nodes, a gate vector $\boldsymbol{\delta}_k \in \mathbb{R}^{|\mathcal{V}_{k}|}$ is introduced. 
After applying a sigmoid activation, a field-wise gating vector 
$\boldsymbol{g}_k = \sigma(\boldsymbol{\delta}_k)$ is obtained, 
where each element $g_v$ corresponds to the importance weight of node $v$. 
The gated node representation is then computed as
\begin{equation}
    \tilde{\mathbf{h}}_v^{(0)} = g_v \cdot \mathbf{h}_v^{(0)}. 
\end{equation}
The sigmoid function constrains the gating weights to the range $(0,1)$, enabling soft feature selection while maintaining differentiability. 
This mechanism enables the model to adaptively emphasize semantically informative protocol fields and provides a natural basis for feature-level interpretability. 

\paragraph{Structure-aware Message Passing}
After feature alignment and gating, each layer expert performs structure-aware message passing on the PTG using a two-layer graph attention architecture. 

The first graph attention layer is designed to capture diverse latent semantics from protocol fields through multi-head attention. 
For a target node $v \in \mathcal{V}_{k}$, its representation after the first attention layer is computed by aggregating messages from its neighbors: 
\begin{align}
    \mathbf{h}_v^{(1)} 
    &= \mathop{\Big\|}_{h=1}^{H} \sum_{u \in \mathcal{N}(v)} \alpha_{vu}^{(h)} \mathbf{W}^{(h)} \tilde{\mathbf{h}}_u^{(0)}, 
\end{align}
where $\alpha_{vu}^{(h)}$ denotes the attention coefficient of the $h$-th head, $\mathbf{W}^{(h)}$ is the corresponding linear projection matrix, and $\Vert$ represents feature concatenation across $H$ attention heads. 
This multi-head mechanism allows the model to attend to protocol semantics from multiple representation subspaces simultaneously. 

The second graph attention layer adopts a single-head attention scheme to further integrate the multi-head representations produced by the first layer. 
This layer focuses on unifying latent semantic information within the expert, yielding a coherent layer-level representation. 
Formally, the output of the second attention layer is given by 
\begin{align}
    \mathbf{h}_v^{(2)} 
    &= \sum_{u \in \mathcal{N}(v)} \alpha_{vu} \mathbf{W} \mathbf{h}_u^{(1)},
\end{align}
where a single attention head is used to aggregate and refine the representations learned from the previous layer.

Through this hierarchical attention-based message passing process, each layer expert effectively captures both local protocol field interactions and global structural semantics encoded in the protocol tree.

\paragraph{Graph Readout}

To obtain a fixed-dimensional representation for each protocol layer, a graph-level readout operation is applied after the final message passing layer. 
Specifically, node representations are aggregated using a global pooling function: 
\begin{equation}
    \mathbf{z}_k = \operatorname{GlobalPool}\left( \left\{ \mathbf{h}_v^{(2)} \mid v \in \mathcal{V}_k \right\} \right),
\end{equation}
where $\mathbf{z}_k \in \mathbb{R}^{D}$ denotes the semantic embedding of the $k$-th protocol layer, and $\operatorname{GlobalPool}(\cdot)$ represents a global mean operation. 
The resulting layer-level embedding $\mathbf{z}_k$ serves as the output of the corresponding protocol layer expert and will be further processed by the mixture-of-experts fusion module. 

\subsection{Optional Flow-Level Expert} \label{subsec: flow_expert}

In addition to protocol-layer experts operating on packet-level protocol tree graphs, PTGAMoE optionally incorporates a flow-level expert to capture global traffic characteristics that are not explicitly represented at the packet field level. 
Flow-level features summarize the holistic behavior of a traffic conversation, such as temporal patterns and statistical distributions, and therefore provide complementary information to protocol tree representations. 

Let $\mathbf{x}_f \in \mathbb{R}^{F}$ denote the aggregated flow-level feature vector associated with a traffic conversation, where $F$ is the number of extracted flow statistics. 
The flow-level expert employs a lightweight feed-forward network to project $\mathbf{x}_f$ into the shared latent space: 
\begin{equation}
    \mathbf{z}_f = f_{\mathrm{flow}}(\mathbf{x}_f),
\end{equation}
where $\mathbf{z}_f \in \mathbb{R}^{D}$ is the resulting flow-level embedding, whose dimensionality is compatible with the fusion stage. 

Unlike protocol-layer experts, which perform structure-aware message passing on protocol tree graphs, the flow-level expert operates on pre-aggregated statistics and captures coarse-grained traffic dynamics. 
When enabled, the flow-level embedding $\mathbf{z}_f$ is treated as an additional expert output and is jointly fused with protocol-layer embeddings in the subsequent mixture-of-experts fusion stage. 
When flow-level features are unavailable or intentionally excluded, the flow-level expert can be omitted without affecting the overall architecture. 
The flow features we used are displayed in the APPENDIX~\ref{apdx: flow_feature} section. 

\subsection{Mixture-of-Experts Fusion}

PTGAMoE employs a mixture-of-experts fusion mechanism to adaptively integrate heterogeneous semantic information from different protocol layers and the flow-level expert when enabled. 
This design allows the model to dynamically adjust the contribution of each expert according to the input traffic characteristics. 

Let $\{\mathbf{z}_1, \mathbf{z}_2, \dots, \mathbf{z}_K\}$ denote the output embeddings of the $K$ protocol-layer experts, where $\mathbf{z}_k \in \mathbb{R}^{D}$ represents the semantic embedding of the $k$-th protocol layer. 
When the optional flow-level expert is enabled, an additional embedding $\mathbf{z}_f \in \mathbb{R}^{D}$ is included. 
In this situation, we denote the complete set of expert embeddings as $\{\mathbf{z}_k\}_{k=1}^{K'}$, where $K' = K + \mathbb{I}_f$. 
$\mathbb{I}_f$ is a indicator function. 
When the flow expert is enabled, $\mathbb{I}_f = 1$, otherwise $\mathbb{I}_f = 0$. 

To enable input-conditioned expert selection, PTGAMoE adopts a lightweight gating network that dynamically computes expert importance for each traffic sample. 
Specifically, all expert embeddings are first concatenated to form a joint representation:
\begin{equation}
    \mathbf{z}_{\mathrm{concat}} = \mathop{\Big\|}_{k=1}^{K'} \mathbf{z}_k.
\end{equation}

A gating function $g(\cdot)$ parameterized by a multi-layer perceptron is then applied to produce expert-level gating scores:
\begin{equation}
    \boldsymbol{\gamma} = g(\mathbf{z}_{\mathrm{concat}}),
\end{equation}
where $\boldsymbol{\gamma} \in \mathbb{R}^{K'}$ depends on the current input sample.

Instead of enforcing competitive normalization, we employ a sigmoid activation to obtain cooperative expert weights:
\begin{equation}
    \boldsymbol{\omega} = \sigma(\boldsymbol{\gamma}),
\end{equation}
where each $\omega_k \in (0,1)$ represents the input-dependent contribution of the $k$-th expert. 
This cooperative formulation allows multiple layer experts to contribute simultaneously, reflecting the complementary nature of hierarchical protocol semantics. 
Unlike competitive softmax routing, which enforces mutual exclusion among experts, sigmoid-based gating enables flexible multi-expert collaboration without constraining the weights to sum to one. 

The final fused representation is computed as a weighted aggregation of expert embeddings:
\begin{equation} \label{eq:pkt_embedding}
    \mathbf{z} = \sum_{k=1}^{K'} \omega_k \cdot \mathbf{z}_k.
\end{equation}
The aggregated representation $\mathbf{z}$ is fed into a feed-forward prediction head to produce packet-level class logits:
\begin{equation} \label{eq:pkt_logits}
    \hat{\mathbf{y}}^{(\mathrm{pkt})} = f_{\mathrm{pred}}^{(\mathrm{pkt})}(\mathbf{z}),
\end{equation}
where $f_{\mathrm{pred}}^{(\mathrm{pkt})}(\cdot)$ denotes a multi-layer perceptron with learnable parameters for packet-level prediction. 

\subsection{Flow-Level Aggregation \& Prediction} 

Confront with the strict scenario in traffic classification tasks, PTGAMoE introduces a flow-level aggregation mechanism that integrates packet-level representations within the same traffic flow to produce a unified flow representation. 

Let a traffic flow $\mathcal{F} = \{ p_{i} \}_{i=1}^{N_{f}}$, where $N_{f}$ denotes the number of packets associated with the flow. 
For each packet $p_{i}$, the model produces a fused representation $\mathbf{z}_{i}$ and corresponding packet-level logits $\hat{\mathbf{y}_{i}}$ according to equation (\ref{eq:pkt_embedding})
and (\ref{eq:pkt_logits}). 

\paragraph{Packet Representation Projection}
The packet-level representations are projected into a latent space suitable for flow aggregation: 
\begin{equation}
    \mathbf{h}_{i} = f_{\mathrm{agg}}(\mathbf{z}_i), 
\end{equation}
where $f_{\mathrm{agg}}(\cdot)$ is implemented as a linear project layer with layer normalization, and $\mathbf{h}_{i} \in \mathbb{R}^{D'}$ denotes the projected packet embedding. 

\paragraph{Flow-Level Representation Aggregation}
A permutation-invariant aggregation function is applied to obtain a flow-level represetation: 
\begin{equation}
    \mathbf{h}^{(\mathcal{F})} = \operatorname{MaxPool}\left( \left\{ \mathbf{h}_i \mid p_{i} \in \mathcal{F} \right\} \right).  
\end{equation}
Here $\operatorname{MaxPool}(\cdot)$ represents the max-pooling operation, which is used to emphasize the most informative packet-level signals within each flow by taking the maximum element-wise packet representations. 
In practice, to ensure computational efficiency and stable training, the aggregation is performed over at most the first $N_{p}$ packets associated with each flow under a micro-macro batching strategy\cite{MicroBatch}, which is illustrated in the APPENDIX~\ref{apdx: batch_settings} section. 

\paragraph{Flow-Level Prediction}
The aggregated flow-level representation $\mathbf{h}^{(\mathcal{F})}$ is then passed through a prediction head to produce flow-level logits: 
\begin{equation}
    \hat{\mathbf{y}}^{(\mathrm{flow})} = f_{\mathrm{pred}}^{(\mathrm{flow})}(\mathbf{h}^{(\mathcal{F})}),
\end{equation}
where $\hat{\mathbf{y}}$ denotes the predicted class logits for the entire traffic flow. 

Given the ground-truth flow label $\mathbf{y}$, the main classification loss is defined as 
\begin{equation}
    \mathcal{L}^{(\mathrm{flow})} = \mathcal{L}_{f}(\hat{\mathbf{y}}^{(\mathrm{pkt})}, \mathbf{y}), 
\end{equation}
where $\mathcal{L}_{f}$ denotes the Focal Loss. 
To stabilize optimization and preserve packet-level discriminative capability, an auxiliary packet-level loss is additionally introduced: 
\begin{equation}
    \mathcal{L}^{(\mathrm{pkt})} = \frac{1}{N}\sum_{i=1}^{N}\mathcal{L}_{f}(\hat{\mathbf{y}}^{(\mathrm{pkt})}, \mathbf{y}). 
\end{equation}
Thus, the final training objective is defined as a weighted combination: 
\begin{equation}
    \mathcal{L} = \mathcal{L}^{(\mathrm{flow})} + \lambda_{1}\mathcal{L}^{(\mathrm{pkt})} + \lambda_{2}\sum_{k} \mathcal{H}(\boldsymbol{g}_k), 
\end{equation}
where $\sum_{k}\mathcal{H}(\boldsymbol{g}_k)$ denotes the feature gating entropy for field-wise gating vector $\boldsymbol{g}_k$. 
$\lambda_{1}$, set to $0.3$ in our experiments, controls the auxiliary packet-level supervision. 
$\lambda_{2}$ is a small regularization coefficient (e.g., $10^{-4}$) that encourages moderate differentiation among field-level gating values. 

\section{Evaluation}

\subsection{Experimental Settings}

\paragraph{Datasets}
To evaluate the performance of PTGAMoE under a no-leakage experimental scenario, we employ two benchmark datasets covering modern TLS1.3 traffic scenarios. 

\begin{itemize}
    \item \textbf{CSTNET-TLS1.3} is a public benchmark dataset for TLS1.3-encrypted web traffic classification, consisting of exclusively encrypted sessions across 26 domains without unencrypted traffic components. 
    \item \textbf{CipherSpectrum} is a contemporary public dataset designed for cipher-agnostic encrypted traffic classification tasks, containing 120,000 TLS1.3-encrypted sessions across 41 domains with uniform coverage of three mandated/recommended TLS1.3 cipher suites (1,000 sessions per suite per class). 
\end{itemize}


\paragraph{Baselines}
To ensure a fair and comprehensive evaluation, we select three of the most popular state-of-the-art (SOTA) open-source models. 
Specifically, ET-BERT, YaTC and RBLJAN are chosen to be the baseline SOTA models, representing tokenized pre-training, image-like matrix processing, and byte-level methods, respectively. 
To rigorously assess genuine classification capabilities, model performance is evaluated under a strict setting where explicit Strong Identification Information (SII) is excluded. 
This includes Ethernet layer attributes, source/destination IP addresses, source/destination port numbers, and the Server Name Indication (SNI) field within the TLS handshake. 
Notably, to construct a data-leakage-free experimental scenario, all flows sharing the same 5-tuple are assigned to the same subset, meaning that flows are strictly isolated across the training, validation, and test sets. 

\paragraph{Metrics}
In addition to standard classification metrics derived from the confusion matrix (e.g., macro-F1), we introduce gate-based metrics to analyze the importance and utilization patterns of protocol fields and layer experts in PTGAMoE.

\begin{itemize}
\item \textbf{Normalized Gate Importance (NGI).}
PTGAMoE employs sigmoid-based gating at both field and expert levels. 
For field-level analysis, given gate values $\{g_i\}$ with $g_i \in (0,1)$, NGI is defined using temperature-scaled softmax normalization:
\begin{equation} \label{eq: ngi_field}
    \tilde{g}_{i} =
    \frac{\exp(g_i / \tau)}{\sum_j \exp(g_j / \tau)},
\end{equation}
where $\tau$ controls the sharpness of the distribution and is set to $0.2$.
For expert-level analysis, let $\omega_k(\mathbf{x})$ denote the input-conditioned expert weight for sample $\mathbf{x}$. 
We first compute the expected expert importance
\begin{equation} 
    \bar{\omega}_k = \mathbb{E}_{\mathbf{x}}[\omega_k(\mathbf{x})],
\end{equation}
and apply the same normalization:
\begin{equation}
    \tilde{\omega}_k =
    \frac{\exp(\bar{\omega}_k / \tau)}{\sum_j \exp(\bar{\omega}_j / \tau)}.
\end{equation}
NGI therefore provides normalized importance scores that enable direct comparison across protocol fields or experts.

\item \textbf{Gate Concentration Ratio (GCR).}
To measure how concentrated the model's preference is, we define the Gate Concentration Score as the Shannon entropy of the NGI distribution:
\begin{equation} \label{eq: gcs}
    \mathrm{GCS} = -\sum_i \tilde{g}_i \log \tilde{g}_i .
\end{equation}
Lower GCS indicates that the model focuses on a small subset of dominant fields or experts, while higher GCS implies more distributed utilization.
For a layer or expert containing $N$ gated elements, we additionally examine the Gate Concentration Ratio 
\begin{equation} \label{eq:gcr}
    \mathrm{GCR} = \frac{log N}{\mathrm{GCS}} 
\end{equation}
to characterize how close the distribution is to uniform. 
When this ratio approaches $1$, the NGI distribution is close to uniform, indicating balanced utilization. 
As the ratio deviates from $1$ (e.g., $\log N / \mathrm{GCS} > 1$), the model exhibits increasingly selective preference toward a subset of features or experts. 
\end{itemize}

\paragraph{Experimental Environment}
We implement the PTGAMoE prototype on a server with an AMD Ryzen 5 5600G CPU, 32\,GB  memory, an NVIDIA GeForce RTX 4060 GPU (8\,GB), and Ubuntu 22.04. 
The software environment is configured with Python 3.12.8 and PyTorch 2.5.1, with CUDA 12.6 enabled for GPU acceleration. 
Graph-based operations are implemented using the PyTorch Geometric library (version 2.6.1). 

\subsection{Classification Performance}

\begin{figure}[htbp]
    \centering
    \includegraphics[width=\linewidth]{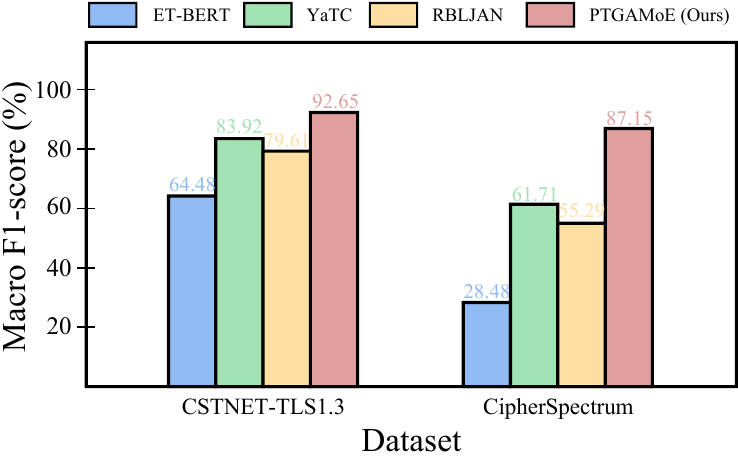}
    \caption{Macro-F1 performance comparison under strict scenarios. }
    \label{fig:models_performance}
\end{figure}

We evaluate PTGAMoE against three representative SOTA models, namely ET-BERT, YaTC, and RBLJAN, on two benchmark datasets under strict no-leakage and weak-indicator settings with flow expert disabled. 
The results are summarized in Fig.~\ref{fig:models_performance}. 

On CSTNET-TLS1.3, PTGAMoE achieves a macro-F1 score of 92.65\%, significantly outperforming all baselines. 
In comparison, RBLJAN and YaTC achieve 83.92\% and 79.61\%, respectively, while ET-BERT reaches only 64.48\%. 
This demonstrates that PTGAMoE can effectively capture protocol semantics even when strong identification features are removed. 
On CipherSpectrum, PTGAMoE also achieves the best performance with a macro-F1 of 87.15\%, substantially surpassing RBLJAN, YaTC, and ET-BERT. 
The performance gap is particularly pronounced compared to ET-BERT, whose reliance on sequential patterns becomes insufficient under strict feature constraints. 

Overall, PTGAMoE consistently outperforms all baselines across datasets, with improvements of over 8.7\% on CSTNET-TLS1.3 and 25.4\% on CipherSpectrum compared to the strongest baseline. 
These results indicate that PTGAMoE provides a more robust and generalizable solution for encrypted traffic classification under realistic deployment conditions. 

\subsection{Field-level Interpretability Analysis}

To investigate the decision mechanism of PTGAMoE under strict no-leakage and flow-isolated settings, we analyze field importance using Normalized Gate Importance (NGI) referring to equation (\ref{eq: ngi_field}) and Gate Concentration Ratio (GCR) referring to equation (\ref{eq:gcr}).

\begin{figure*}[htbp]
    \centering    
    \includegraphics[width=\linewidth]{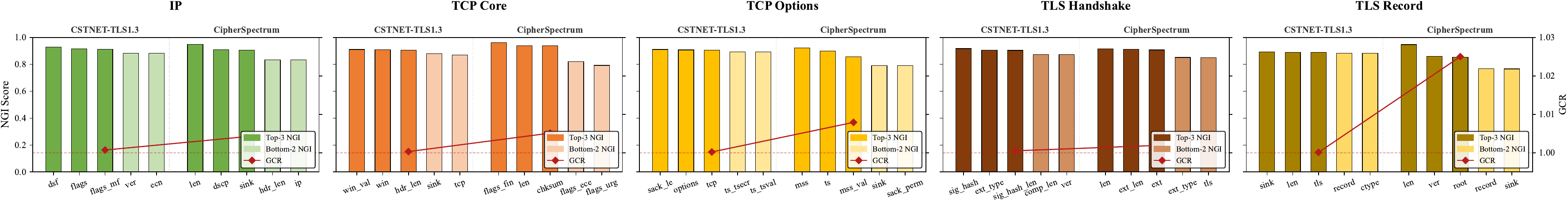}
    \caption{Field-level importance (NGI) and concentration (GCR) across protocol layers. }
    \label{fig: field-level NGI}
\end{figure*}

As shown in Fig.~\ref{fig: field-level NGI}, consistent patterns emerge across datasets. 
At the IP layer, traffic-handling fields such as differentiated services fields and flags are consistently ranked higher than static identifiers (e.g., version-related fields), indicating reliance on packet processing semantics. 
At the TCP core layer, fields related to flow control and connection state (e.g., window size, header length, and flags) dominate, reflecting connection-level behavior rather than identifier-based cues. 
For TCP options, structural fields (e.g., SACK, MSS) are generally more important than timestamp-related features, suggesting that temporal signals are not universally dominant. 
At higher layers, the TLS handshake focuses on negotiation semantics (e.g., signature/hash algorithms and extensions), while the TLS record layer is primarily driven by structural features such as length and aggregation fields. 
Notably, the sink node appears as a top feature only in certain cases, indicating that it acts as a conditional global aggregator rather than a shortcut feature. 

Compared to conventional packet-level models, which often rely on strong identifiers (e.g., IP, ports, or SNI), PTGAMoE demonstrates a clear shift toward behavior-driven features, including packet size patterns, transport dynamics, and protocol structure. 
This confirms that the model captures intrinsic protocol semantics under strict settings. 

The GCR analysis further shows that feature importance remains well-distributed within each layer. 
On CSTNET-TLS1.3, all layers exhibit GCR values close to 1, indicating highly balanced feature utilization. 
On CipherSpectrum, slightly higher GCR values are observed, peaking at 1.025 particularly in the TLS record layer, which suggests a moderate dataset-specific concentration on structural features. 

Overall, these results demonstrate that PTGAMoE performs semantics-aware and distributed reasoning, effectively avoiding shortcut learning while adaptively leveraging both local protocol fields and global context.

\subsection{Layer-level Interpretability Analysis}

\begin{figure}[htbp]
    \centering    
    \includegraphics[width=\linewidth]{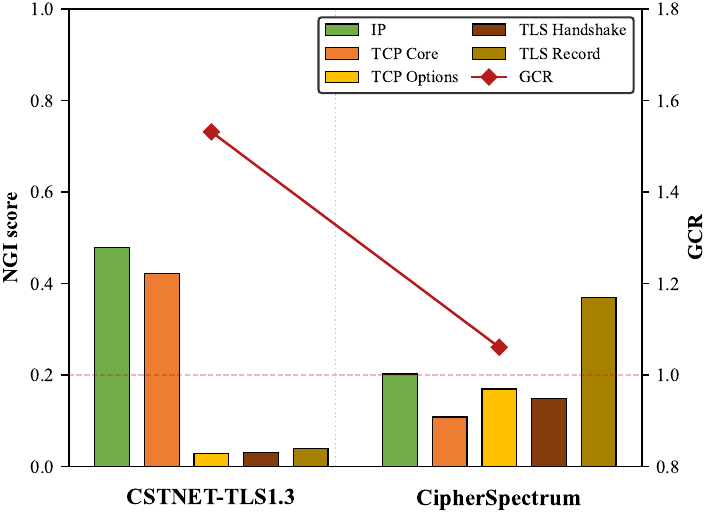}
    \caption{Layer-level expert importance and contribution patterns. }
    \label{fig: level-level NGI}
\end{figure}

We further analyze the layer-level expert behavior using NGI and GCR to understand how PTGAMoE allocates importance across protocol layers. 

On CSTNET-TLS1.3, the model exhibits strong layer concentration, with IP and TCP core dominating the decision process, while higher-layer experts contribute marginally. 
This is reflected by a high GCR value of 1.53, indicating that classification primarily relies on transport-level behavior. 
This suggests that discriminative patterns in this dataset are mainly captured by packet handling and connection dynamics rather than TLS semantics. 
In contrast, CipherSpectrum shows a much more balanced expert distribution, where all experts contribute evenly to yield a GCR of 1.06. 
The TLS record expert becomes the most important component, indicating that structural features at the application layer play a dominant role. 
Meanwhile, IP, TCP options, and TLS handshake also contribute significantly, demonstrating cross-layer collaboration. 

These results reveal that PTGAMoE adaptively adjusts expert importance according to dataset characteristics, while avoiding expert collapse. 
The model selectively emphasizes the most informative protocol layers while maintaining multi-layer semantic integration, confirming its ability to capture intrinsic traffic behavior under strict settings. 

\subsection{Impact of Strong Indication Information}

\begin{figure*}[htbp]
    \centering    
    \includegraphics[width=\linewidth]{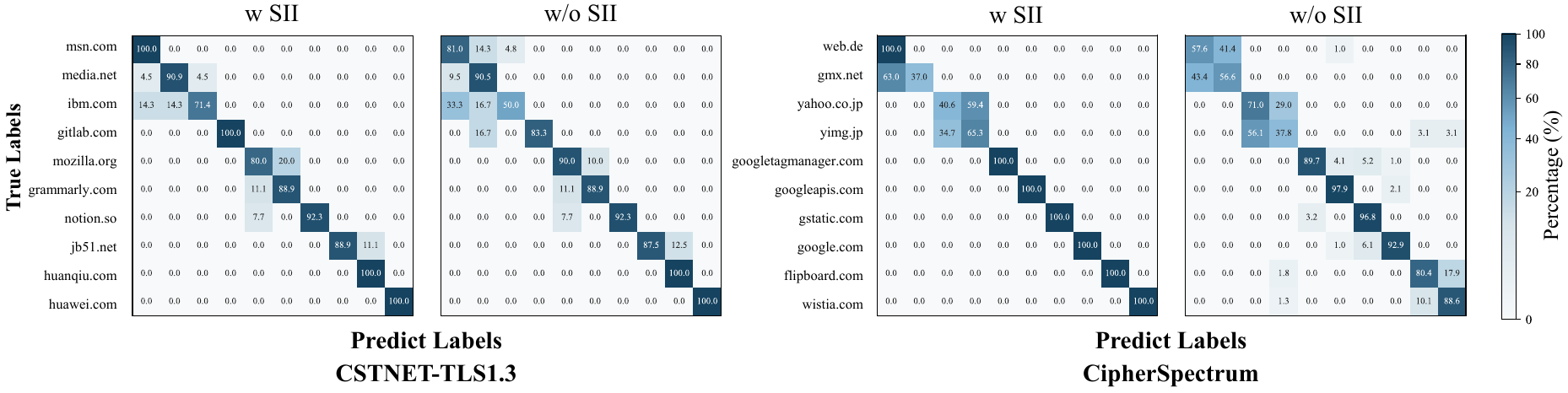}
    \caption{Impact of Strong Identification Information (SII) on classification confusion patterns. }
    \label{fig:cm_sii}
\end{figure*}

To assess whether the performance gain mainly comes from intrinsic protocol semantics or shortcut identity cues, we further compare PTGAMoE under settings with and without strong indication information (SII). 
The macro-F1 results are reported in Table~\ref{tab:ablation_analysis}, while representative confusion matrices are shown in Fig.~\ref{fig:cm_sii}.

\begin{table}[t]
\centering
\caption{Ablation results for SII and Flow Expert (Macro-F1 \%).}
\label{tab:ablation_analysis}
\footnotesize
\setlength{\tabcolsep}{4pt}
\begin{tabular}{lcc}
\toprule
Setting 
& CSTNET-TLS1.3 
& CipherSpectrum \\
\midrule
\textbf{Final (w/o SII, w/o Flow Expert)} 
& \textbf{92.65}
& \textbf{87.15} \\
\midrule
w SII 
& 93.05 
& 94.23 \\
$\Delta_{\text{SII}}$ 
& +0.40 
& +7.08 \\
\midrule
w Flow Expert 
& 92.55 
& 47.27 \\
$\Delta_{\text{Flow}}$ 
& -0.10 
& -39.88 \\
\bottomrule
\end{tabular}
\end{table}

As shown in Table~\ref{tab:ablation_analysis}, the impact of SII is highly dataset-dependent. 
On CSTNET-TLS1.3, the macro-F1 score increases only marginally from 92.65\% to 93.05\%, indicating that most discriminative information can already be captured from protocol structure, transport behavior, and flow-level dynamics. 
In contrast, on CipherSpectrum, the macro-F1 score rises substantially from 87.15\% to 94.23\%, suggesting that this dataset contains much stronger identity-related cues and is therefore more sensitive to shortcut features. 

The representative confusion matrices of CSTNET-TLS1.3 further support this observation. 
Even without SII, most selected domains remain well separated, and the residual errors are concentrated in a small number of confusing groups, most notably \texttt{msn.com}, \texttt{media.net}, and \texttt{ibm.com}. 
Additional minor confusions can also be observed around \texttt{gitlab.com}, \texttt{mozilla.org}, \texttt{grammarly.com}, and \texttt{notion.so}. 
After introducing SII, these residual errors are only slightly reduced, which is consistent with the marginal overall F1 improvement. 
This indicates that CSTNET-TLS1.3 is primarily semantics-dominant, where classification mainly relies on intrinsic protocol behavior rather than strong identity indicators. 

By contrast, the representative confusion matrices of CipherSpectrum shows a much heavier dependency on SII. 
Without SII, substantial confusion appears among domains sharing similar infrastructure, service ecosystems, or content delivery patterns. 
Typical examples include the portal pair \texttt{web.de} and \texttt{gmx.net}, the regional pair \texttt{yahoo.co.jp} and \texttt{yimg.jp}, as well as several Google-related services such as \texttt{googletagmanager.com}, \texttt{googleapis.com}, \texttt{gstatic.com}, and \texttt{google.com}. 
These confusions are significantly alleviated once SII is introduced, leading to a much cleaner class separation. 
This behavior indicates that CipherSpectrum is considerably more SII-sensitive, and that strong indicators can simplify the task by providing direct identity cues. 

Overall, these results show that the availability of SII can substantially alter task difficulty and evaluation credibility. 
In particular, the large gain on CipherSpectrum suggests that retaining strong indicators may overestimate model capability by allowing shortcut learning. 
Therefore, removing SII is necessary for rigorously evaluating whether a model truly learns behavior-driven protocol semantics rather than superficial identifiers. 
The relatively small performance drop on CSTNET-TLS1.3, together with the still strong performance on CipherSpectrum under the w/o SII setting, demonstrates that PTGAMoE remains effective under strict weak-indicator conditions. 

\subsection{Impact of the Flow Expert} 

\begin{figure*}[htbp]
    \centering    
    \includegraphics[width=\linewidth]{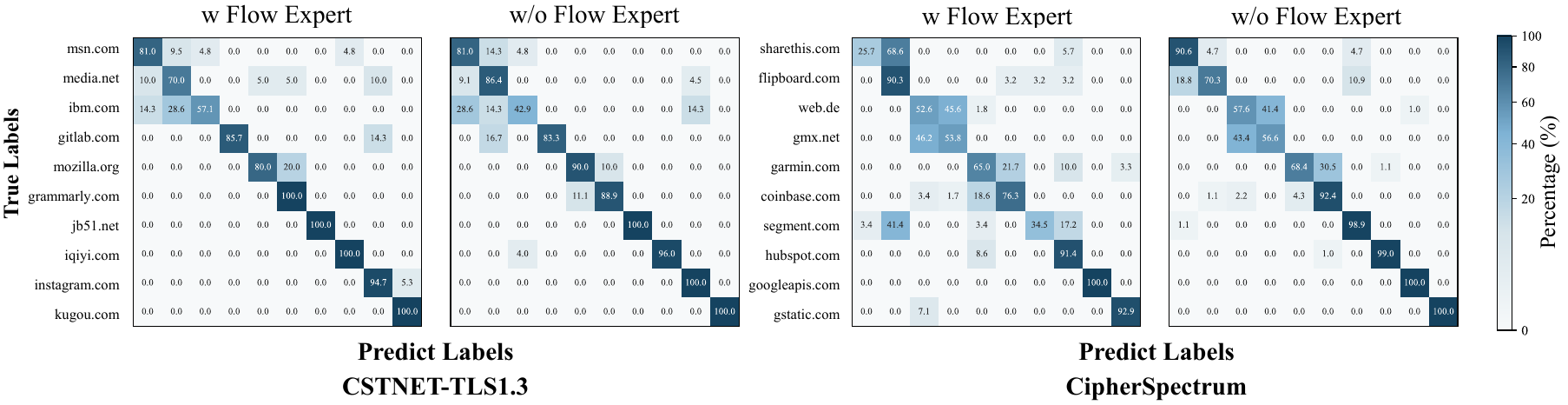}
    \caption{Confusion matrices comparison between settings with and without the Flow Expert. }
    \label{fig:cm_flow}
\end{figure*}

We further evaluate whether an explicit flow-level expert provides additional benefits beyond the proposed flow-level aggregation mechanism. 

As shown in TABLE~\ref{tab:ablation_analysis}, the impact of the Flow Expert is highly negative overall. 
On CSTNET-TLS1.3, the macro-F1 score changes only marginally from 92.65\% to 92.55\%, indicating that the Flow Expert provides no meaningful improvement. 
On CipherSpectrum, the macro-F1 score drops drastically from 87.15\% to 47.27\%, revealing a severe performance degradation. 

As shown in Fig.~\ref{fig:cm_flow}, the representative confusion matrices further illustrate this difference. 
On CSTNET-TLS1.3, the error patterns remain largely similar with and without the Flow Expert, and only minor fluctuations are observed on a few classes. 
This suggests that the Layer Expert Committee and the flow-level aggregation mechanism already capture most of the useful discriminative information. 
However, the Flow Expert causes widespread confusion across many classes on CipherSpectrum. 
Without the Flow Expert, the confusion is relatively structured and primarily occurs within semantically related groups, such as the portal pair \texttt{web.de} and \texttt{gmx.net}, or clusters of tracking and content delivery services (e.g., \texttt{sharethis.com}, \texttt{flipboard.com}, and \texttt{segment.com}). 
However, after introducing the Flow Expert, the confusion becomes widespread and no longer constrained within these semantic clusters. 
Many classes that were previously well distinguished, such as \texttt{sharethis.com}, \texttt{coinbase.com}, and \texttt{hubspot.com}, exhibit severe degradation, with predictions spreading across a large number of unrelated classes. 
This indicates that the Flow Expert introduces a strong noise source that disrupts the learned decision boundaries. 

These results suggest that the performance gain of the proposed flow-enhanced framework mainly comes from flow-level aggregation and supervision, rather than from an explicit Flow Expert based on coarse-grained statistical features. 
In complex multi-class scenarios such as CipherSpectrum, the Flow Expert may even become misleading, as many classes share similar flow statistics while differing in finer protocol semantics. 
Therefore, we exclude the Flow Expert in the final configuration and retain only the flow-level aggregation mechanism. 

\section{Conclusion}

In this paper, we present PTGAMoE, a protocol-aware framework for encrypted traffic analysis that explicitly models the hierarchical structure of network protocols. 
By integrating Protocol Tree Graph Attention (PTGA) with a layer-wise Mixture-of-Experts (MoE) architecture, the model captures fine-grained field semantics without disruptive padding. 
Crucially, we introduce a permutation-invariant flow aggregation mechanism to ensure robust classification within strict scenarios, where the exclusion of strong identification information (SII) and the implementation of flow-isolated splitting prevent the exploitation of deceptive shortcuts. 
Extensive experiments on modern TLS 1.3 datasets demonstrate that PTGAMoE consistently outperforms SOTA models.
Furthermore, hierarchical gating mechanisms and the proposed NGI and GCS metrics provide quantifiable interpretability at both field and protocol levels. 
These results reinforce that protocol-native structural modeling provides a reliable, semantic-preserving foundation for robust encrypted traffic analysis. 
Future work will explore extending PTGAMoE to a broader range of protocols (e.g., UDP-based protocols and proxy-related protocols) and more diverse network environments. 

\bibliographystyle{IEEEtran}  
\bibliography{references}            

@article{backgroud_survey1,
  title={A survey on encrypted network traffic analysis applications, techniques, and countermeasures},
  author={Papadogiannaki, Eva and Ioannidis, Sotiris},
  journal={ACM Computing Surveys (CSUR)},
  volume={54},
  number={6},
  pages={1--35},
  year={2021},
  publisher={ACM New York, NY, USA}
}

@article{sota-tdsc25-xiao,
  title={RBLJAN: Robust Byte-Label Joint Attention Network for Network Traffic Classification},
  author={Xiao, Xi and Wang, Shuo and Hu, Guangwu and Li, Qing and Mao, Kelong and Luo, Xiapu and Zhang, Bin and Xia, Shutao},
  journal={IEEE Transactions on Dependable and Secure Computing},
  year={2024},
  publisher={IEEE}
}

@inproceedings{sota-aaai25-qiu,
  title={Dual-Channel Interactive Graph Transformer for Traffic Classification with Message-Aware Flow Representation},
  author={Qiu, Xing and Cheng, Guang and Zhu, Weizhou and Niu, Dandan and Fu, Nan},
  booktitle={Proceedings of the AAAI Conference on Artificial Intelligence},
  volume={39},
  number={1},
  pages={685--693},
  year={2025}
}

@article{MoE1_KIMI,
  title={Kimi k2: Open agentic intelligence},
  author={Team, Kimi and Bai, Yifan and Bao, Yiping and Chen, Guanduo and Chen, Jiahao and Chen, Ningxin and Chen, Ruijue and Chen, Yanru and Chen, Yuankun and Chen, Yutian and others},
  journal={arXiv preprint arXiv:2507.20534},
  year={2025}
}

@article{MoE2_Google,
  title={Scaling vision with sparse mixture of experts},
  author={Riquelme, Carlos and Puigcerver, Joan and Mustafa, Basil and Neumann, Maxim and Jenatton, Rodolphe and Susano Pinto, Andr{\'e} and Keysers, Daniel and Houlsby, Neil},
  journal={Advances in Neural Information Processing Systems},
  volume={34},
  pages={8583--8595},
  year={2021}
}

@article{MoE3_Xiaomi,
  title={MiMo: Unlocking the Reasoning Potential of Language Model--From Pretraining to Posttraining},
  author={Xiaomi, LLM and Xia, Bingquan and Shen, Bowen and Zhu, Dawei and Zhang, Di and Wang, Gang and Zhang, Hailin and Liu, Huaqiu and Xiao, Jiebao and Dong, Jinhao and others},
  journal={arXiv preprint arXiv:2505.07608},
  year={2025}
}

@article{MoE4_Qwen,
  title={Qwen3-vl technical report},
  author={Bai, Shuai and Cai, Yuxuan and Chen, Ruizhe and Chen, Keqin and Chen, Xionghui and Cheng, Zesen and Deng, Lianghao and Ding, Wei and Gao, Chang and Ge, Chunjiang and others},
  journal={arXiv preprint arXiv:2511.21631},
  year={2025}
}

@article{MoE5_survey,
  title={A survey on mixture of experts in large language models},
  author={Cai, Weilin and Jiang, Juyong and Wang, Fan and Tang, Jing and Kim, Sunghun and Huang, Jiayi},
  journal={IEEE Transactions on Knowledge and Data Engineering},
  year={2025},
  publisher={IEEE}
}

@inproceedings{Intro_TLS1,
  title={A comprehensive symbolic analysis of TLS 1.3},
  author={Cremers, Cas and Horvat, Marko and Hoyland, Jonathan and Scott, Sam and Van Der Merwe, Thyla},
  booktitle={Proceedings of the 2017 ACM SIGSAC conference on computer and communications security},
  pages={1773--1788},
  year={2017}
}

@inproceedings{Intro_TLS2,
  title={A symbolic analysis of privacy for tls 1.3 with encrypted client hello},
  author={Bhargavan, Karthikeyan and Cheval, Vincent and Wood, Christopher},
  booktitle={Proceedings of the 2022 ACM SIGSAC Conference on Computer and Communications Security},
  pages={365--379},
  year={2022}
}

@inproceedings{RW_representation1_Kitsune,
  title={Kitsune: An Ensemble of Autoencoders for Online Network Intrusion Detection},
  author={Mirsky, Yisroel and Doitshman, Tomer and Elovici, Yuval and Shabtai, Asaf},
  booktitle={25th Annual Network and Distributed System Security Symposium, NDSS 2018},
  year={2018},
  organization={The Internet Society}
}

@inproceedings{RW_representation2_TrafficFormer,
  title={Trafficformer: an efficient pre-trained model for traffic data},
  author={Zhou, Guangmeng and Guo, Xiongwen and Liu, Zhuotao and Li, Tong and Li, Qi and Xu, Ke},
  booktitle={2025 IEEE symposium on security and privacy (SP)},
  pages={1844--1860},
  year={2025},
  organization={IEEE}
}

@inproceedings{RW_representation3_ETBERT,
  title={Et-bert: A contextualized datagram representation with pre-training transformers for encrypted traffic classification},
  author={Lin, Xinjie and Xiong, Gang and Gou, Gaopeng and Li, Zhen and Shi, Junzheng and Yu, Jing},
  booktitle={Proceedings of the ACM Web Conference 2022},
  pages={633--642},
  year={2022}
}

@inproceedings{RW_representation4_Yet,
  title={Yet another traffic classifier: A masked autoencoder based traffic transformer with multi-level flow representation},
  author={Zhao, Ruijie and Zhan, Mingwei and Deng, Xianwen and Wang, Yanhao and Wang, Yijun and Gui, Guan and Xue, Zhi},
  booktitle={Proceedings of the AAAI Conference on Artificial Intelligence},
  volume={37},
  number={4},
  pages={5420--5427},
  year={2023}
}

@inproceedings{RW_representation5_PktSemantics,
  title={Detecting tunneled flooding traffic via deep semantic analysis of packet length patterns},
  author={Fu, Chuanpu and Li, Qi and Shen, Meng and Xu, Ke},
  booktitle={Proceedings of the 2024 on ACM SIGSAC Conference on Computer and Communications Security},
  pages={3659--3673},
  year={2024}
}

@inproceedings{RW_representation6_NetTraceAug,
  title={Realistic website fingerprinting by augmenting network traces},
  author={Bahramali, Alireza and Bozorgi, Ardavan and Houmansadr, Amir},
  booktitle={Proceedings of the 2023 ACM SIGSAC Conference on Computer and Communications Security},
  pages={1035--1049},
  year={2023}
}

@inproceedings{RW_representation7_FS-NET,
  title={Fs-net: A flow sequence network for encrypted traffic classification},
  author={Liu, Chang and He, Longtao and Xiong, Gang and Cao, Zigang and Li, Zhen},
  booktitle={IEEE INFOCOM 2019-IEEE Conference On Computer Communications},
  pages={1171--1179},
  year={2019},
  organization={IEEE}
}

@inproceedings{RW_representation8_YouDPI,
  title={You do (not) belong here: detecting DPI evasion attacks with context learning},
  author={Zhu, Shitong and Li, Shasha and Wang, Zhongjie and Chen, Xun and Qian, Zhiyun and Krishnamurthy, Srikanth V and Chan, Kevin S and Swami, Ananthram},
  booktitle={Proceedings of the 16th International Conference on emerging Networking EXperiments and Technologies},
  pages={183--197},
  year={2020}
}

@article{RW_representation9_TimeTrantra,
  title={Tantra: Timing-based adversarial network traffic reshaping attack},
  author={Sharon, Yam and Berend, David and Liu, Yang and Shabtai, Asaf and Elovici, Yuval},
  journal={IEEE Transactions on Information Forensics and Security},
  volume={17},
  pages={3225--3237},
  year={2022},
  publisher={IEEE}
}

@article{RW_gnn1_DGNN,
  title={DGNN: Accurate Darknet Application Classification Adopting Attention Graph Neural Network},
  author={Zhu, Yuehao and Tao, Jun and Wang, Haotian and Yu, LinXiao and Luo, Yuantu and Qi, Tianyi and Wang, Zuyan and Xu, Yifan},
  journal={IEEE Transactions on Network and Service Management},
  year={2023},
  publisher={IEEE}
}

@article{RW_gnn2_FlowGNN,
  title={Flow-based encrypted network traffic classification with graph neural networks},
  author={Huoh, Ting-Li and Luo, Yan and Li, Peilong and Zhang, Tong},
  journal={IEEE Transactions on Network and Service Management},
  volume={20},
  number={2},
  pages={1224--1237},
  year={2022},
  publisher={IEEE}
}

@article{RW_gnn3_GSPB,
  title={GSPB: a global-statistic and packet-byte fusion framework for encrypted traffic classification},
  author={Li, Haiyue and Tao, Jun and Yu, Linxiao and Luo, Yuantu and Wang, Zuyan},
  journal={Cybersecurity},
  volume={8},
  number={1},
  pages={120},
  year={2025},
  publisher={Springer}
}

@inproceedings{RW_gnn4_AGMF,
  title={Attention-Guided Multi-view Feature Fusion for Proxy Traffic Classification},
  author={Tang, Xu and Tao, Jun and Luo, Yuantu},
  booktitle={International Conference on Neural Information Processing},
  pages={425--439},
  year={2025},
  organization={Springer}
}

@article{RW_gnn5_flow,
  title={Flow-based encrypted network traffic classification with graph neural networks},
  author={Huoh, Ting-Li and Luo, Yan and Li, Peilong and Zhang, Tong},
  journal={IEEE Transactions on Network and Service Management},
  volume={20},
  number={2},
  pages={1224--1237},
  year={2022},
  publisher={IEEE}
}

@article{RW_moe1_CL-ViME,
  title={CL-ViME: Contrastive Learning and Vision Mixture of Experts for Encrypted Traffic Classification},
  author={Cai, Saihua and Chen, Lizhou and Chen, Jinfu and Wang, Shengran and Zhang, Guofeng},
  journal={IEEE Transactions on Network and Service Management},
  volume={23},
  pages={1422--1434},
  year={2025},
  publisher={IEEE}
}

@article{RW_moe2_SwitchTransformer,
  title={Switch transformers: Scaling to trillion parameter models with simple and efficient sparsity},
  author={Fedus, William and Zoph, Barret and Shazeer, Noam},
  journal={Journal of Machine Learning Research},
  volume={23},
  number={120},
  pages={1--39},
  year={2022}
}

@inproceedings{RW_moe3_Time-MoE,
  title={Time-MoE: Billion-Scale Time Series Foundation Models with Mixture of Experts},
  author={Xiaoming, Shi and Shiyu, Wang and Yuqi, Nie and Dianqi, Li and Zhou, Ye and Qingsong, Wen and Jin, Ming},
  booktitle={ICLR 2025: The Thirteenth International Conference on Learning Representations},
  year={2025},
  organization={International Conference on Learning Representations}
}

@inproceedings{RW_moe4_Moirai-MoE,
  title={Moirai-MoE: Empowering Time Series Foundation Models with Sparse Mixture of Experts},
  author={Liu, Xu and Liu, Juncheng and Woo, Gerald and Aksu, Taha and Liang, Yuxuan and Zimmermann, Roger and Liu, Chenghao and Li, Junnan and Savarese, Silvio and Xiong, Caiming and others},
  booktitle={International Conference on Machine Learning},
  pages={38940--38962},
  year={2025},
  organization={PMLR}
}

@inproceedings{RW_sok_enigma,
  title={Sok: Decoding the enigma of encrypted network traffic classifiers},
  author={Wickramasinghe, Nimesha and Shaghaghi, Arash and Tsudik, Gene and Jha, Sanjay},
  booktitle={2025 IEEE Symposium on Security and Privacy (SP)},
  pages={1825--1843},
  year={2025},
  organization={IEEE}
}

@inproceedings{RW_sok_sweet,
  title={The sweet danger of sugar: Debunking representation learning for encrypted traffic classification},
  author={Zhao, Yuqi and Dettori, Giovanni and Boffa, Matteo and Vassio, Luca and Mellia, Marco},
  booktitle={Proceedings of the ACM SIGCOMM 2025 Conference},
  pages={296--310},
  year={2025}
}

@misc{dissection_tree,
author = {Gerald Combs},
  title = {Adding Information To The Dissection Tree},
  xhowpublished = {\url{https://www.wireshark.org/docs/}},
  url = {https://www.wireshark.org/docs//wsdg_html_chunked/lua_module_Tree.html},
  note = {Accessed: 2025-12-20},
}

@misc{RFC_TLS1.3,
    series =    {Request for Comments},
    number =    8446,
    howpublished =  {RFC 8446},
    publisher = {RFC Editor},
    doi =       {10.17487/RFC8446},
    url =       {https://www.rfc-editor.org/info/rfc8446},
    author =    {Eric Rescorla},
    title =     {{The Transport Layer Security (TLS) Protocol Version 1.3}},
    pagetotal = 160,
    year =      2018,
    month =     aug,
}

@article{Method_GCN-RTG,
  title={Accurate compressed traffic detection via traffic analysis using Graph Convolutional Network based on graph structure feature},
  author={Fu, Nan and Cheng, Guang and Su, Xinyue},
  journal={Computer Communications},
  volume={207},
  pages={128--139},
  year={2023},
  publisher={Elsevier}
}

@inproceedings{Method_GAT,
  title={Graph Attention Networks},
  author={Veli{\v{c}}kovi{\'c}, Petar and Cucurull, Guillem and Casanova, Arantxa and Romero, Adriana and Li{\`o}, Pietro and Bengio, Yoshua},
  booktitle={International Conference on Learning Representations},
  year={2018}
}

@inproceedings{MicroBatch,
  title={Accelerating distributed dlrm training with optimized tt decomposition and micro-batching},
  author={Wang, Weihu and Xia, Yaqi and Yang, Donglin and Zhou, Xiaobo and Cheng, Dazhao},
  booktitle={SC24: International Conference for High Performance Computing, Networking, Storage and Analysis},
  pages={1--15},
  year={2024},
  organization={IEEE}
}


\appendices
\section{Implementation Details}

\subsection{Streaming Field Extraction in Field-Level Preprocessing} \label{apdx: streaming_field_extraction} 

Network traffic can be captured by tools like Wireshark, which saves the captured traffic packets into PCAP files. 
Due to the long time development and modification of global researchers and developers, Wireshark has abundant inner dissectors for various network protocols. 
The dissectors dissect the raw packet bytes by the structural protocol architectures and display the dissected packet fields into a dissection tree. 
The dissection tree not only can be examined in Wireshark GUI, but also can be exported into a detailed PDML (Packet Description Markup Language) file, which conforms to the XML standard and contains details about the packet dissection. 
Leveraging the PDML file, we can extract the hierarchical fields into the most common CSV format. 

However, the PDML file is a double-edged sword, which makes the direct extraction cause huge memory and storage expenditure, because of its global detailed field records for the whole PCAP file. 
A MB-level PCAP may generate a GB-level XML, which is unacceptable in the view of memory and storage. 
To address this issue, we design a streaming PCAP-to-CSV converter that couples tShark’s byte pipe PDML output with chunk-wise constant memory parsing, addressing the bottleneck of traditional offline processing. 
Each chunk is immediately appended to the target CSV, eliminating GB level temporary files and bounding RAM consumption to $O(c)$, in which $c$ is the size of each chunk. 
Complexity drops from $\Omega(n)$ memory of the classic PDML to CSV pipeline to $O(c)$ memory, yielding about $10 \times$ peak RAM reduction.  

\subsection{Batch Settings under the Strict Scenario}\label{apdx: batch_settings}

To support strict flow-isolated training and maximize GPU utilization, we employ a hierarchical micro-macro batching mechanism. 

\paragraph{Micro-Batch for Flow-centric Truncation}
To maintain flow-level structural integrity, each individual flow with $N_{f}$ packets $\mathcal{F}_i = \{p_{i,1}, \dots, p_{i,N_{f}}\}$ is treated as a single micro-batch. 
We define a micro-batch size $N_p$ (max packets per flow) to truncate or pad the sequences: 
\begin{equation}
    \widetilde{\mathcal{F}}_i = \{p_{i,1}, \dots, p_{i,\min(N_{f}, N_p)}\}.
\end{equation}
This ensures that all packets within a micro-batch belong to the same flow, preventing cross-flow interference and facilitating the permutation-invariant aggregation stage. 

\paragraph{Macro-Batch for Parallel GPU Acceleration}
For computational efficiency, multiple micro-batches are aggregated into a macro-batch $\mathcal{B}_t$ for each optimization step $\mathcal{S}_t$. 
Given a macro-batch size $K_f$ (flows per step), the batch is constructed as: 
\begin{equation}
    \mathcal{B}_t = \bigcup_{i \in \mathcal{S}_t} \widetilde{\mathcal{F}}_i, \quad |\mathcal{S}_t| = K_f. 
\end{equation}
This design allows the GPU to process $K_f$ flows in parallel while keeping the total packet count per step bounded by $|\mathcal{B}_t|\leq K_f N_p$. 

In our implementation, we set $N_p=64$ and $K_f=64$.

\subsection{Flow-Level Feature Engineering} \label{apdx: flow_feature}

When enabled, the flow-level expert utilizes eight statistical features computed by grouping packets according to transport-layer identifiers. 
To prevent information leakage, all statistics are computed independently for each data split. 
The features include: (1) count (total packets). 
(2) length (mean, std. dev., and ratio of packets $>1400$ bytes). 
(3) IAT (mean, std. dev., and max inter-arrival time). 
(4) normalized duration (flow duration per packet). 
Most features are normalized by packet count to ensure consistency across datasets, and timing-related metrics are only extracted when valid timestamps are available. 




\vspace{11pt}



\vfill

\end{document}